%% file: bus.tex
\documentstyle[twocolumn,aps]{revtex}
\begin{document}

\include{incpic}

\emptyplace{3.3in \includegraphics{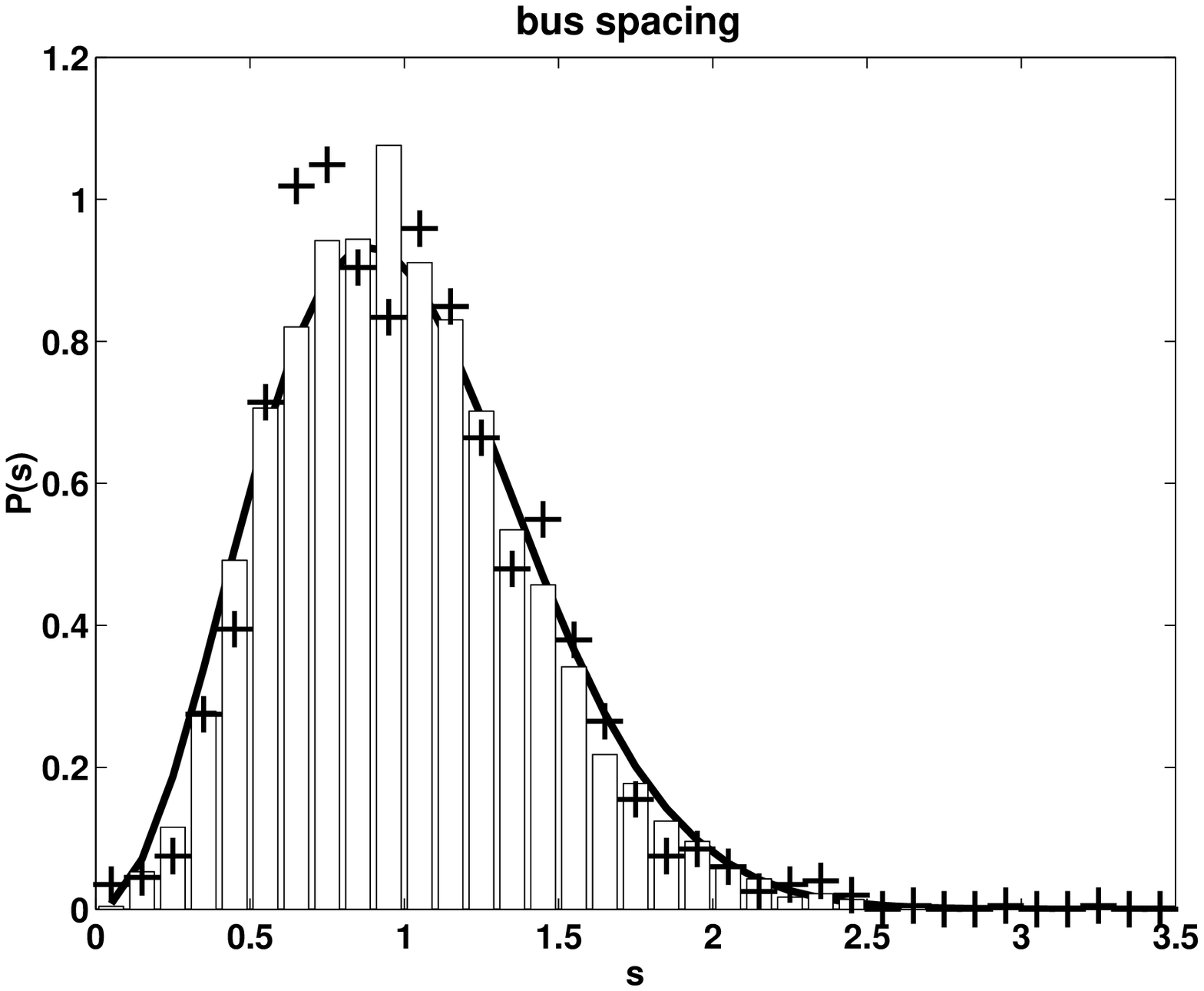}} {3.1in} {\footnotesize
\noindent FIG.1. Bus interval distribution $P(s)$ obtained for the
city line No. 4. The full line represents the random matrix
prediction (\ref{spacing}), the markers (+) represent the bus
interval data and bars display the random matrix prediction
(\ref{spacing}) with $0.8 \%$ of the data rejected }

\emptyplace{3.3in \includegraphics{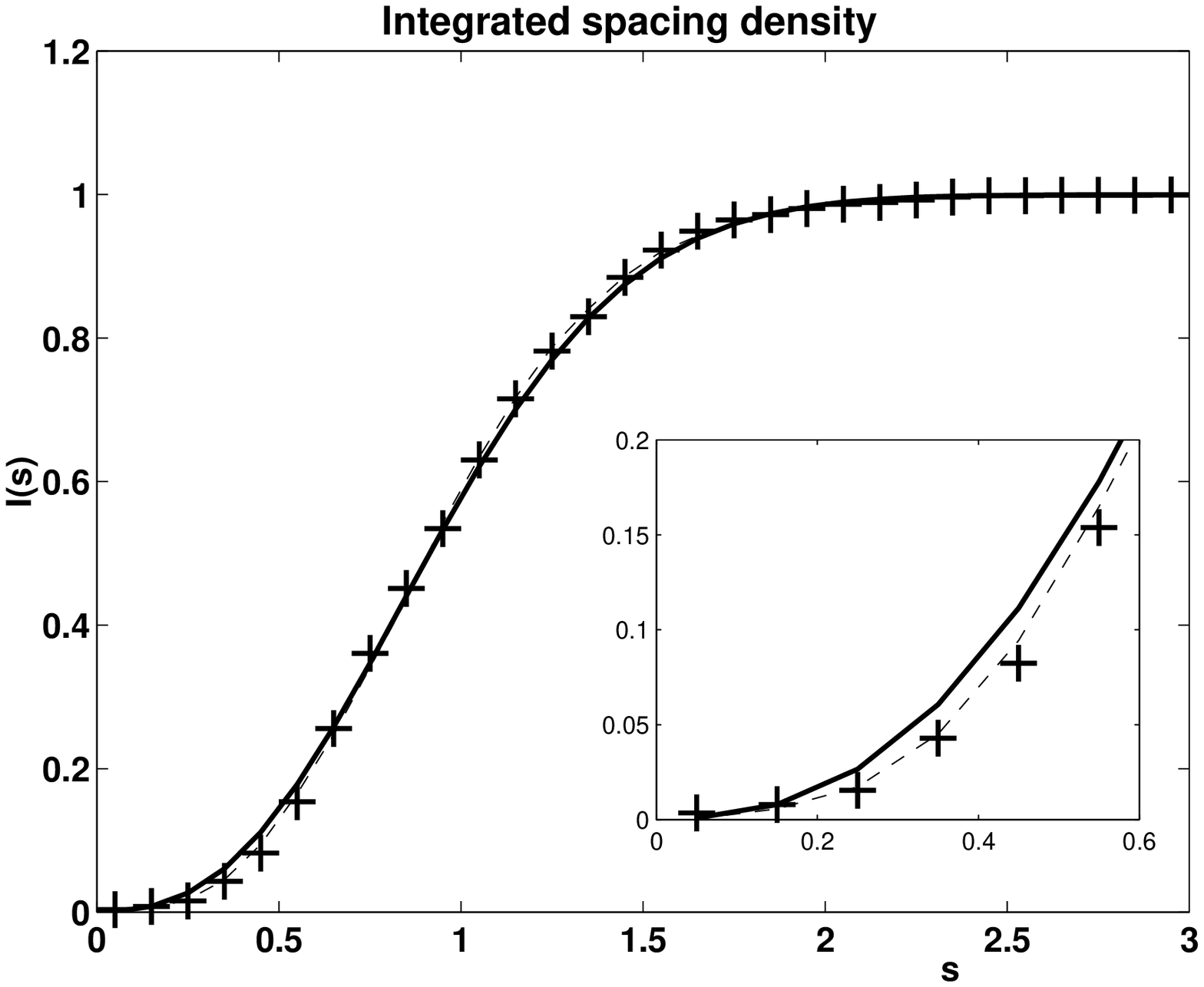}} {3.1in} {\footnotesize
\noindent FIG.2. The integrated bus interval distribution $I(s)$
The full line represents the random matrix prediction, the markers
(+) represent the bus data and dashed line shows the random matrix
prediction  with $0.8 \%$ of the data rejected. The distribution
close to the origin is magnified on the insert }

\emptyplace{3.3in \includegraphics{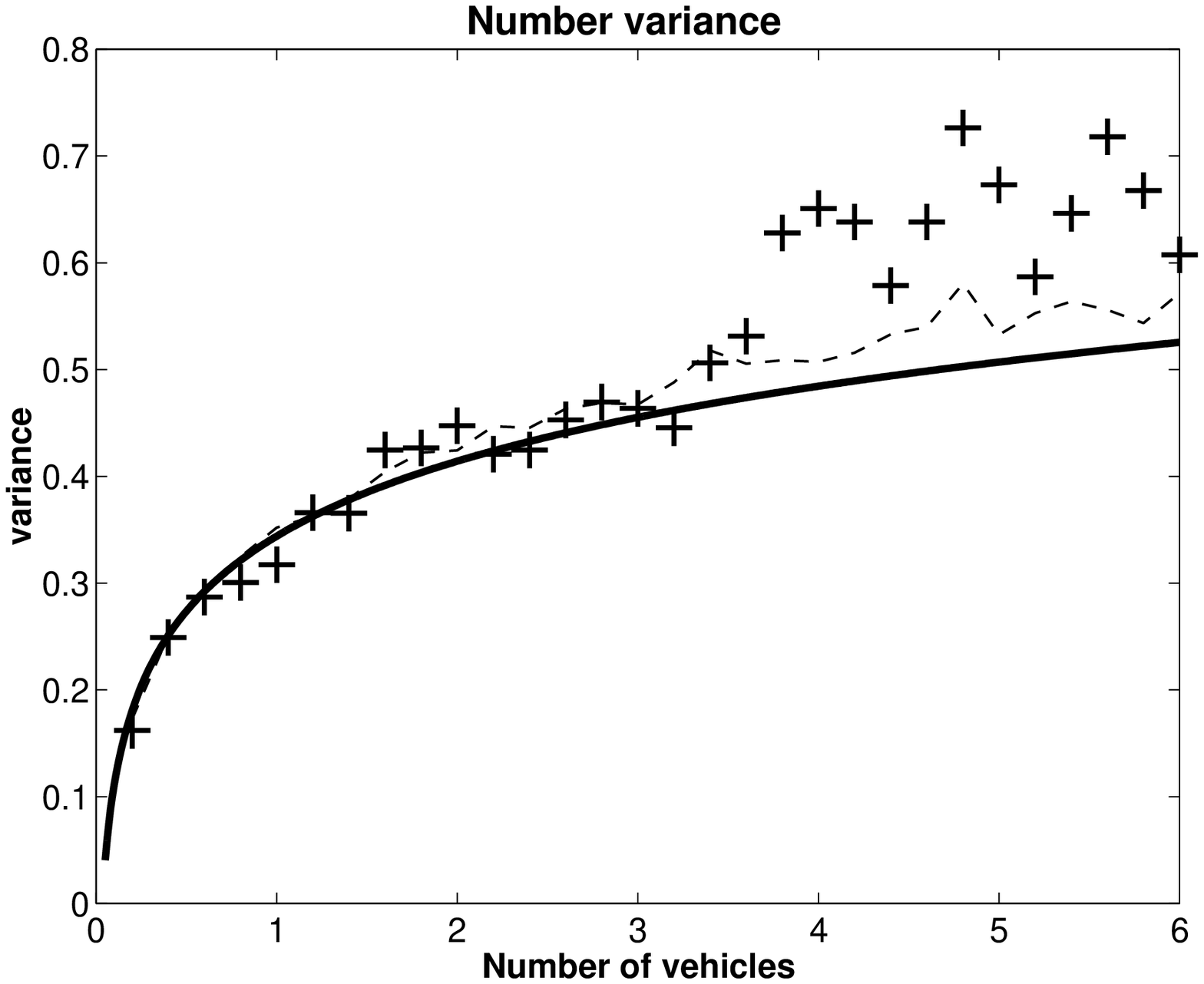}} {3.1in} {\footnotesize
\noindent FIG.3. The number variance $N(s)$. Full line represents
the random matrix prediction (\ref{variance}), the markers (+)
represent the bus data and the dashed line display result obtained
with the help of the potential (\ref{dyson}) whith the summation
restricted to three neighboring particles only. }

\twocolumn[\hsize\textwidth\columnwidth\hsize\csname@twocolumnfalse\endcsname

\title{The  statistical properties of the city transport in Cuernavaca (Mexico) and Random matrix ensembles}

\author{Milan Krb\' alek ${ }^{1,3}$ and  Petr \v Seba ${ }^{2,3}$}
\address{
 ${ }^1$ \ Department of Mathematics, Faculty of Nuclear Sciences and Physical Engineering,
  Trojanova 13, Prague
 Czech republic\\
 ${ }^2$ \ Institute of Physics, Czech Academy of Sciences,
Cukrovarnicka 10, Prague, Czech Republic\\
 ${ }^3$ \ Department of
Physics, Pedagogical University, Vita Nejedleho 573, Hradec
Kralove, Czech Republic\\ }

\maketitle
%
%\widetext
%
\begin{abstract}
We analyze  statistical properties of the city bus transport in
Cuernavaca (Mexico) and show that the bus arrivals display
probability distributions  conforming those given by the Unitary
Ensemble of random matrices.
\end{abstract}

\pacs{05.45+b, 05.20-y, 05.60+w}
\vskip2pc]

 \narrowtext

%\begin{twocolumn}

It is well known that the statistical properties of coherent
chaotic quantum systems are well described by the Wigner/ Dyson
random matrix ensembles. The fact that the spectral statistics of
such chaotic systems is to a large extend generic  - a phenomenon
known as the universality of quantum chaos -  has been confirmed
both theoretically and experimentally.(See for instance
\cite{Haake} for references.)

The statistical distributions characterizing  the ensembles of
random matrices can be understood as minimizing the information
contained in the system with the constrains that the matrices
posses some discrete symmetry properties \cite{MeBa},
\cite{MePeSe}. Let $P(x_1,x_2,...,x_n)$ denotes the joint
probability distribution of the eigenvalues $x_1,x_2,...,x_n$ of
the given matrices and
\begin{equation}
I=-\int P(x_1,x_2,...,x_n)\ln(P(x_1,x_2,...,x_n)) dx_1...dx_n
\label{info}
\end{equation}
be its information content. Assuming for instance that the
matrices are invariant with respect to a time reversal
transformation the information $I$ is minimized when the
distribution $P(x_1,x_2,...,x_n)$ describes Orthogonal ensemble
(GOE). If there is not external symmetry the total minimum of the
information $I$ is achieved for the Unitary ensemble (GUE), where
the only constrain is that the matrices should be hermitean.

It is known for a long time  that  matrix ensembles are of
relevance also for classical one dimensional interacting many
particle systems, where the matrix eigenvalues $x_1,x_2,...,x_n$
describe the positions of the particles. So the thermal
equilibrium of a one-dimensional gas interacting via Coulomb
potential (Dyson gas) has statistical properties (depending on
temperature) that are identical with those of random matrix
ensembles \cite{Mehta}. The same holds true also for other
potentials. An example is the Pechukas gas \cite{Pechukas} where
the one dimensional particles interact by a potential $\lambda
V(x)$ with $V(x)= 1/\vert x\vert ^2$,  $x$ being their mutual
distance and $\lambda$ the relevant coupling constant. Regarding
the couplings $\lambda$ as additive canonical variables it has
been shown by Pechukas \cite{Pechukas} and Yukawa \cite{Yukawa},
that the statistical equilibrium of the related canonical ensemble
is  described by  random matrix theory. It has to be stressed
however, that those results were obtained under special
requirements on the dynamics of the variables $\lambda$, ensuring
in fact full equivalence of the system to matrix diagonalization.
Nevertheless the methods of statistical physics remain valid also
for different shapes of the particle potential as well as for
different dynamics of the coupling variables $\lambda$. It can be
shown that the potential
\begin{equation}
V(x)\approx 1/\vert x\vert ^a
\label{potential}
\end{equation}
with $a$ being positive constant, leads also to random matrix
distribution of the particle positions. The equivalence of the
statistical properties of the particle positions  of  one
dimensional interacting gases to random matrix ensembles and the
fact that GUE minimizes the information (\ref{info}) lead us to
speculate, that whenever the information contained in the gas is
minimized its properties are described by GUE. However, according
to our best knowledge, this fact was never tested.

The one dimensional gas to be studied in the present letter is
represented by buses that operate the city line number 4 in
Cuernavaca (Mexico). We will show  that the statistical properties
of the bus arrivals are described by the Unitary Ensemble of
random matrices. To explain the origin of the interaction between
subsequent buses several remarks are necessary . First of all it
has to be stressed that there is not a covering company
responsible for organizing the city transport. Consequently such
constrains like a time table etc. that represent external
influence on the transport do not exist. Moreover, each bus is a
property of the driver. The drivers try to maximize their income
and hence the  amount of passengers they transport. This lead to
competition among the drivers and to their mutual interaction. It
is clear that without interaction the probability distribution of
the distances between subsequent buses will be Poissonian. (This
is due to the rather complicated traffic conditions in the city
that work as an effective randomizer). Poisson distribution imply,
however, that the probability of close encounters of two buses is
high (bus clustering) which is in conflict with the effort of the
driver to maximize the number of transported passengers and
accordingly maximize the distance to the preceding  bus. In order
to avoid the unpleasant clustering effect the bus drivers in
Cuernavaca  engage people that record the arrival times of buses
on significant places . Arriving at a checkpoint, the driver gets
the information when the previous bus passed that place. Knowing
the time interval to the preceding bus  the driver tries to
optimize the distance to it either by  slowing down or speeding
up. In such a way the obtained information leads to interaction
between buses and changes their statistical properties.

We have collected  records catching the arrivals of the buses of
the line No.4  close to the city center. The record contains
altogether 3500 arrivals during a time period of 27 days whereby
the arrivals on different days are regarded  as statistically
independent.  After unfolding the peak times we evaluated the
related probability distributions and compared them with the
predictions of GUE.   In particular we have focused on the bus
spacing distribution, i.e. on the probability density $P(s)$ that
spacing between two subsequent buses equals to $s$ and on the bus
number variance N(T) measuring the fluctuations of the total
number $n(T)$ of buses arriving to the place during the time
interval $T$:
\begin{equation}
N(T)=<\left (n(T)-T\right )^2>
\end{equation}
where $< >$ means the sample averaging. (Note that after unfolding
the mean distance between buses equals to 1.) According to the
prediction of the unitary ensemble the spacing distribution and
the number variance are given by
\begin{equation}
P(s)= \frac{32}{\pi^2} s^2 \exp{\left (-\frac{4}{\pi} s^2\right )}
\label{spacing}
\end{equation}
and
\begin{equation}
N(T)\approx \frac{1}{\pi^2}\left(\ln{2 \pi T} + \gamma +1 \right)
\label{variance}
\end{equation}

Those predictions are compared with the obtained bus arrival data
and displayed on the following figures:

Figure 1 shows the bus interval distribution when compared with
the GUE prediction (\ref{spacing}). The bus data are marked by
$(+)$. The minor discrepancy between the GUE prediction and the
bus data can be explained taking into account the fact that the
bus data do not represent the full record. Assuming that roughly
$0.8 \%$ of the bus arrivals is not notified and rejecting the
same amount of randomly chosen data from the random matrix
eigenvalues, we get very satisfactory agreement.
 Due to the limited amount of records available the bus
interval distribution is sensitive to the  binning used in the
evaluation of the probability density $P(s)$.
 \inspicture

This is why  on the next figure we plot the integrated interval
distribution $I(s)=\int P(s')ds'$ that is not liable to binning
fluctuations. The agreement with the GUE distribution is evident.

\inspicture

The next figure shows the number variance (\ref{variance})
obtained for GUE and compared with the bus data. Here the
agreement is good up to time interval $T \approx 3$. For larger
$T$ the number variance of the bus arrivals lies significantly
above the prediction given by (\ref{variance}). This indicates
that the long range correlations between more then three buses are
weaker then predicted by the Unitary ensemble. The explanation is
simple: getting the time interval information of the preceding bus
the driver tries to optimize his position . Doing so he has,
however, to take into account also the assumed interval to the bus
behind him since otherwise this bus will overtake him. Hence the
driver tries to optimize his position between the preceding and
following bus that leads to the observed correlation.

 \inspicture

The GUE properties of the bus arrival statistics can be understood
when regarding the buses as one dimensional interacting gas. It
was already mentioned that the exact GUE statistics is obtained
for Coulomb interaction between the gas particles, i.e. for the
interaction potential $V$ given by
\begin{equation}
V=-\sum_{i<j} log(\vert x_i - x_j \vert )+\frac{1}{2}\sum_i x_i^2
\label{dyson}
\end{equation}
(In (\ref{dyson})  the second terms represents a force confining
the gas close to origin and is not important for our discussion.
Equivalently one can discuss a one dimensional gas on a circle
instead and then the second term is missing). The statistical
properties of the particle positions of  the Dyson gas
 are identical with those of the random
matrix ensembles \cite{Dyson},\cite{Scharf}. In particular the
properties of the unitary ensemble are recovered by minimizing the
information contained in the particle positions.

It is of interest that similar potential can be indeed found when
studying the reaction of driver on the traffic situation. Here we
can use older results describing the behaviour of highway drivers.
For one dimensional  models it was shown \cite{Montroll} that the
i-th driver accelerate according to
\begin{equation}
\frac{dv_i}{dt}\approx \frac{f(v_{i+1},v_{i})}{x_{i+1}-x{i}}
\label{trafic}
\end{equation}
where $x_{i+1}$ and $v_{i+1}$ represent the position and velocity
of the preceding car respectively and $f(v_{i+1},v_{i})$ is a
function depending on the car velocities only. Approximating $f$
by a constant (justified for low velocities) we get that the cars
accelerate in the same way as described by  the Coulomb
interaction (\ref{dyson}).

The exact form of the potential is, however,  not crucial for the
result. Using Metropolis algorithm \cite{Scharf} we have
numerically evaluated the equilibrium distributions of the
positions  of  one dimensional gas interacting via potential
(\ref{potential}). When the exponent $a$ is fixed and $a<2$ the
resulting equilibrium distributions belong to the same class as in
the Dyson case (\ref{dyson}). The numerical results show clearly
that for a given $a$ one can always find such a temperature of the
gas that the equilibrium  distribution is given by GUE. Moreover,
the fact that the original Dyson potential (\ref{dyson}) contains
interaction between all pairs of the gas particles is also not
substantial. Numerical simulations show that a good agreement with
the random matrix theory is obtained when the summation in
(\ref{dyson}) is restricted and involves three neighboring
particles only.

The exact interaction between buses in Cuernavaca is not known.
However the weak sensitivity of the statistical equilibrium to the
exact form of the potential guide us  to the conviction that
unitary ensembles are a good choice for bus  description.

We conclude that  the   statistical properties of the city bus
transport in Cuernavaca can be described by Gaussian Unitary
Ensemble of random matrices. This behavior can be understood as
equilibrium state of interacting one dimensional gas under the
assumption that the information contained in the positions of
individual gas particles is minimized. The agreement of the actual
bus data with the GUE prediction is surprisingly good.

We would like to thank Dr. Markus Mueller from the University in
Cuernavaca who helped us to collect the bus data and to Tomas
Zuscak for patience by entering the collected data to computer.
This work was supported by the Academy of Sciences of the Czech
Republic under Grant No. A1048804 and by the "Foundation for
Theoretical Physics" in Slemeno, Czech Republic.

%\end{twocolumn}

\end{document}

%% file: incpic.tex
% Macro for more comfortable treating pictures (pictures placed
% on left or right side of page are imbedded into paragraphs, pictures
% that are centered can be imbedded between lines of paragraph).
% For more details see the end of this file.
%
% Oldrich Ulrych                                     July 30, 1992, Prague
%
%
\edef\catcodeat{\the\catcode`\@ }     \catcode`\@=11
\newbox\p@ctbox                       % box for the current picture to be placed in
\newbox\t@mpbox                       % box for temporary storing
\newbox\@uxbox                        % auxiliary box
\newbox\s@vebox                       % stack of scanned pictures
\newtoks\desct@ks \desct@ks={}        % stack for of descriptions under pictures
\newtoks\@ppenddesc                   % toks for appended local description
\newtoks\sh@petoks                    % token for description of picture
\newif\ifallfr@med  \allfr@medfalse   % true if all picture are being fr@med
\newif\if@ddedrows                    % true if round brackets appeared
\newif\iffirstp@ss  \firstp@ssfalse   % flag for paragraph test
\newif\if@mbeeded                     % flag to indicate just to copy picture
\newif\ifpr@cisebox                   % flag to copy picture in natural width
\newif\ifvt@p                         % inserted picture is \vtop
\newif\ifvb@t                         % inserted picture is \vbot
\newif\iff@nished    \f@nishedtrue    % text is around the picture
\newif\iffr@med                       % picture will be framed
\newif\ifj@stbox     \j@stboxfalse    % only framed box is shown
\newcount\helpc@unt                   % auxiliary counter
\newcount\p@ctpos                     % position of the picture (0-l,1-c,2-r)
\newdimen\r@leth      \r@leth=0.4pt   % thickness of rules
\newdimen\x@nit                       % x-size of user's horizontal units
\newdimen\y@nit                       % y-size of user's vertical units
\newdimen\xsh@ft                      % x-shift of user's coordinates
\newdimen\ysh@ft                      % y-shift of user's coordinates
\newdimen\@uxdimen                    % auxiliary dimension
\newdimen\t@mpdimen                   % auxiliary dimension
\newdimen\t@mpdimeni                  % auxiliary dimension
\newdimen\b@tweentandp                % marginal space between text and picture
\newdimen\b@ttomedge                  % bottom edge of the picture on page
\newdimen\@pperedge                   % upper edge of the picture on page
\newdimen\@therside                   % space on the outer side of picture
\newdimen\d@scmargin                  % space around marks inside picture
\newdimen\p@ctht                      % height+depth of picture
\newdimen\l@stdepth
\newdimen\in@tdimen
\newdimen\l@nelength
\newdimen\re@lpictwidth
\def\justframes{\global\j@stboxtrue}  % only framed empty place is shown
\def\picturemargins#1#2{\b@tweentandp=#1\@therside=#2\relax}
\def\allframed{\global\allfr@medtrue} % all picturew will be fr@med
\def\emptyplace#1#2{\pl@cedefs        % the empty vbox will be created
    \setbox\@uxbox=\vbox to#2{\n@llpar% all paragraph values are reset
        \hsize=#1\vfil \vrule height0pt width\hsize}% vbox to prescribed dimensions
    \e@tmarks}
\def\boxplace{\pl@cedefs\afterassignment\re@dvbox\let\n@xt= }
\def\re@dvbox{\setbox\@uxbox=\vbox\bgroup% all is stored in vbox
         \n@llpar\aftergroup\e@tmarks}
\def\fontcharplace#1#2{\pl@cedefs     % picture is in font
    \setbox\@uxbox=\hbox{#1\char#2\/}%
    \xsh@ft=-\wd\@uxbox               % width of character and shift of origin
    \setbox\@uxbox=\hbox{#1\char#2}%
    \advance\xsh@ft by \wd\@uxbox     % horizontal shift of the origin
    \helpc@unt=#2
    \advance\helpc@unt by -63         % fontdimen code-63: size of user's x-unit
    \x@nit=\fontdimen\helpc@unt#1%
    \advance\helpc@unt by  20         % fontdimen code-43: size of user's y-unit
    \y@nit=\fontdimen\helpc@unt#1%
    \advance\helpc@unt by  20         % fontdimen code-23: y-shift of origin
    \ifnum\helpc@unt<51
      \ysh@ft=-\fontdimen\helpc@unt#1%
    \fi
    \e@tmarks}
\def\n@llpar{\parskip0pt \parindent0pt% paragraph values are reset
    \leftskip=0pt \rightskip=0pt
    \everypar={}}
\def\pl@cedefs{\xsh@ft=0pt\ysh@ft=0pt}% by default user's origin is not shifted
\def\e@tmarks#1{\setbox\@uxbox=\vbox{ % reset the box with picture
      \n@llpar
      \hsize=\wd\@uxbox               % to preserve its width
      \noindent\copy\@uxbox           % the box itself
      \kern-\wd\@uxbox                % jump to the left margin
      #1\par}%                        % mark into the picture
    \st@redescription}
\def\t@stprevpict#1{\ifvoid#1\else    % if previous picture is not finished
   \errmessage{Previous picture is not finished yet.}\fi} % then error message

\def\st@redescription#1\par{%         % puts picture's description on the stack
    \global\setbox\s@vebox=\vbox{\box\@uxbox\unvbox\s@vebox}%
    \desct@ks=\expandafter{\the\desct@ks#1\@ndtoks}}%remembers the description
\def\def@ultdefs{\p@ctpos=1         % picture is centered by default
      \def\lines@bove{0}%           % no full lines above picture by default
      \@ddedrowsfalse               % no extra full lines read by default
      \@mbeededfalse                % paragraph flows around picture by default
      \pr@ciseboxfalse
      \vt@pfalse                    % output box will be vbox
      \vb@tfalse                    % output box will be vbox
      \@ppenddesc={}%               % no append to description
      \ifallfr@med\fr@medtrue\else\fr@medfalse\fi
      }

\def\descriptionmargins#1{\global\d@scmargin=#1\relax}
\def\@dddimen#1#2{\t@mpdimen=#1\advance\t@mpdimen by#2#1=\t@mpdimen}
\def\placemark#1#2 #3 #4 #5 {\unskip    % ignore previous spaces
      \setbox1=\hbox{\kern\d@scmargin#5\kern\d@scmargin}% form the mark
      \@dddimen{\ht1}\d@scmargin        % add some space above mark
      \@dddimen{\dp1}\d@scmargin        % add some space below mark
      \ifx#1l\dimen3=0pt\else           % the correction
        \ifx#1c\dimen3=-0.5\wd1\else
          \ifx#1r\dimen3=-\wd1
     \fi\fi\fi
     \ifx#2u\dimen4=-\ht1\else          % and @dd the correction
       \ifx#2c\dimen4=-0.5\ht1\advance\dimen4 by 0.5\dp1\else
         \ifx#2b\dimen4=0pt\else
           \ifx#2l\dimen4=\dp1
     \fi\fi\fi\fi
     \advance\dimen3 by #3%             % x position
     \advance\dimen4 by #4%             % y position
     \advance\dimen4 by-\dp1
     \advance\dimen3 by \xsh@ft         % influence of shifting of the origin
     \advance\dimen4 by \ysh@ft         % influence of shifting of the origin
     \kern\dimen3\vbox to 0pt{\vss\copy1\kern\dimen4}% put the mark
     \kern-\wd1                        % jump back of the width of mark
     \kern-\dimen3                     % jump back of the x position
     \ignorespaces}                    % ignore following spaces
\def\fontmark #1#2 #3 #4 #5 {\placemark #1#2 #3\x@nit{} #4\y@nit{} {#5} }
\def\fr@msavetopict{\global\setbox\s@vebox=\vbox{\unvbox\s@vebox
      \global\setbox\p@ctbox=\lastbox}%
    \expandafter\firstt@ks\the\desct@ks\st@ptoks}
\def\firstt@ks#1\@ndtoks#2\st@ptoks{%
    \global\desct@ks={#2}%
    \def\t@mpdef{#1}%
    \@ppenddesc=\expandafter\expandafter\expandafter
                        {\expandafter\t@mpdef\the\@ppenddesc}}
\def\testf@nished{{\let\s@tparshape=\relax
    \s@thangindent}}
\def\inspicture{\t@stprevpict\p@ctbox
    \def@ultdefs                  % set default values
    \fr@msavetopict
    \iff@nished\else\testf@nished\fi
    \iff@nished\else
      \immediate\write16{Previes picture is not finished yet}%
    \fi
    \futurelet\N@xt\t@stoptions}  % look ahead
\def\t@stoptions{\let\n@xt\@neletter% scans parameters
  \ifx\N@xt l\p@ctpos=0\else                % picture on left inside par
   \ifx\N@xt c\p@ctpos=1\else               % picture copied centered
    \ifx\N@xt r\p@ctpos=2\else              % picture on right inside par
     \ifx\N@xt(\let\n@xt\e@tline\else        % number of full row above
      \ifx\N@xt!\@mbeededtrue\else           % just copy the picture
       \ifx\N@xt|\fr@medtrue\else            % picture will be framed
        \ifx\N@xt^\vt@ptrue\vb@tfalse\else  % picture as vtop
         \ifx\N@xt_\vb@ttrue\vt@pfalse\else % picture as vbot
          \ifx\N@xt\bgroup\let\n@xt\@ddgrouptodesc\else
           \let\n@xt\@dddescription % all parameters are read
  \fi\fi\fi\fi\fi\fi\fi\fi\fi\n@xt}
\def\e@tline(#1){\def\lines@bove{#1}% define number of full rows above picture
    \@ddedrowstrue
    \futurelet\N@xt\t@stoptions}
\def\@neletter#1{\futurelet\N@xt\t@stoptions} % eats one parameter
\def\@ddgrouptodesc#1{\@ppenddesc={#1}\futurelet\N@xt\t@stoptions}
\def\fr@medpict{\setbox\p@ctbox=
    \vbox{\n@llpar\hsize=\wd\p@ctbox
       \iffr@med\else\r@leth=0pt\fi
       \ifj@stbox\r@leth=0.4pt\fi
       \hrule height\r@leth \kern-\r@leth
       \vrule height\ht\p@ctbox depth\dp\p@ctbox width\r@leth \kern-\r@leth
       \ifj@stbox\hfill\else\copy\p@ctbox\fi
       \kern-\r@leth\vrule width\r@leth\par
       \kern-\r@leth \hrule height\r@leth}}
\def\@dddescription{\fr@medpict     %  adds description to the picture
    \re@lpictwidth=\the\wd\p@ctbox
    \advance\re@lpictwidth by\@therside
    \advance\re@lpictwidth by\b@tweentandp
    \ifhmode\ifinner\pr@ciseboxtrue\fi\fi
    \createp@ctbox
    \let\N@xt\tr@toplacepicture
    \ifhmode                         % if inside paragraph then use \vadjust
      \ifinner\let\N@xt\justc@py
      \else\let\N@xt\vjustc@py
      \fi
    \else
      \ifnum\p@ctpos=1               % if centered then just copy the box
        \let\N@xt\justc@py
      \fi
    \fi
    \if@mbeeded\let\N@xt\justc@py\fi % if forced to copy, just copy the box
    \firstp@sstrue
    \N@xt}
\def\createp@ctbox{\global\p@ctht=\ht\p@ctbox
    \advance\p@ctht by\dp\p@ctbox
    \advance\p@ctht by 6pt
    \setbox\p@ctbox=\vbox{%        % make new vbox
      \n@llpar                     % reset par's parameters
      \t@mpdimen=\@therside          % and set the leftskip and rightskip
      \t@mpdimeni=\hsize             %  according to the position of picture
      \advance\t@mpdimeni by -\@therside
      \advance\t@mpdimeni by -\wd\p@ctbox
      \ifpr@cisebox
        \hsize=\wd\p@ctbox
      \else
        \ifcase\p@ctpos
               \leftskip=\t@mpdimen    \rightskip=\t@mpdimeni
        \or    \advance\t@mpdimeni by \@therside
               \leftskip=0.5\t@mpdimeni \rightskip=\leftskip
        \or    \leftskip=\t@mpdimeni   \rightskip=\t@mpdimen
        \fi
      \fi
      \hrule height0pt             % invisible rule and nobreak penalty
      \kern6pt                     % small space above
      \penalty10000
      \noindent\copy\p@ctbox\par     % put the picture with marks
      \kern3pt                       % small space between pict and description
      \hrule height0pt
      \hbox{}%
      \penalty10000
      \interlinepenalty=10000
      \the\@ppenddesc\par            % follows text of description
      \penalty10000                  % prohibite page break
      \kern3pt                       % small space below
      }%
      \ifvt@p
       \setbox\p@ctbox=\vtop{\unvbox\p@ctbox}%
      \else
        \ifvb@t\else
          \@uxdimen=\ht\p@ctbox
          \advance\@uxdimen by -\p@ctht
          {\vfuzz=\maxdimen
           \global\setbox\p@ctbox=\vbox to\p@ctht{\unvbox\p@ctbox}%
          }%
          \dp\p@ctbox=\@uxdimen
        \fi
      \fi
      }
\def\picname#1{\unskip\setbox\@uxbox=\hbox{\bf\ignorespaces#1\unskip\ }%
      \hangindent\wd\@uxbox\hangafter1\noindent\box\@uxbox\ignorespaces}
\def\justc@py{\ifinner\box\p@ctbox\else\kern\parskip\unvbox\p@ctbox\fi
  \global\setbox\p@ctbox=\box\voidb@x}
\def\vjustc@py{\vadjust{\kern0.5\baselineskip\unvbox\p@ctbox}%
      \global\setbox\p@ctbox=\box\voidb@x}
\def\tr@toplacepicture{%             % tries to find the good place
      \ifvmode\l@stdepth=\prevdepth  % remember the depth of last box
      \else   \l@stdepth=0pt         % otherwise the depth of last box is zero
      \fi
      \vrule height.85em width0pt\par% to define exact totalpage
      \r@memberdims                  % remembers the significant dimensions
      \global\t@mpdimen=\pagetotal
      \t@stheightofpage              % tests if the picture fits here
      \ifdim\b@ttomedge<\pagegoal    % if the picture ends on the current page
         \let\N@xt\f@gurehere        % the picture can be placed here
         \global\everypar{}%         % clear up the everypar token
      \else
         \let\N@xt\relax             % we must wait till the new page
         \penalty10000
         \vskip-\baselineskip        % but jump back to the reference point
         \vskip-\parskip             % of the last line of the previous par
         \immediate\write16{Picture will be shifted down.}%
         \global\everypar{\sw@tchingpass}%  % check the begin of every paragraph
      \fi
      \penalty10000
      \N@xt}
\def\sw@tchingpass{%                 % on every odd pass t@sts the possibility
    \iffirstp@ss                     %   to place picture here
      \let\n@xt\relax
      \firstp@ssfalse                % on every even pass does nothing
    \else
      \let\n@xt\tr@toplacepicture
      \firstp@sstrue
    \fi  \n@xt}
\def\r@memberdims{\global\in@tdimen=0pt
    \ifnum\p@ctpos=0
        \global\in@tdimen=\re@lpictwidth
      \fi
      \global\l@nelength=\hsize
      \global\advance\l@nelength by-\re@lpictwidth
      }
\def\t@stheightofpage{%
     \global\@pperedge=\t@mpdimen
     \advance\t@mpdimen by-0.7\baselineskip % jump back to the begin of par
     \advance\t@mpdimen by \lines@bove\baselineskip % and down by full lines
     \advance\t@mpdimen by \ht\p@ctbox      % and down by the height of new box
     \advance\t@mpdimen by \dp\p@ctbox      % and down by the depth of new box
     \advance\t@mpdimen by-0.3\baselineskip % and down by the depth of last box
     \global\b@ttomedge=\t@mpdimen          % and remember bottom edge
     }
\def\f@gurehere{\global\f@nishedfalse
      \t@mpdimen=\lines@bove\baselineskip   % and remember the amount of used
      \advance\t@mpdimen-0.7\baselineskip   %   space
      \kern\t@mpdimen
      \advance\t@mpdimen by\ht\p@ctbox
      \advance\t@mpdimen by\dp\p@ctbox
      {\t@mpdimeni=\baselineskip
       \offinterlineskip
       \unvbox\p@ctbox
       \global\setbox\p@ctbox=\box\voidb@x
       \penalty10000   \kern-\t@mpdimen     % and jump back to the reference
       \penalty10000   \vskip-\parskip      % point of the last line of previous
       \kern-\t@mpdimeni                    % paragraph
%       \hbox{\vrule height\l@stdepth width0pt}%
      }%
      \penalty10000                         % prevent page breaking
      \global\everypar{\s@thangindent}%     % and mark that paragraph is special
      }
\def\s@thangindent{%
    \ifdim\pagetotal>\b@ttomedge\global\everypar{}%
      \global\f@nishedtrue             % if we continue under the picture
      \else
        \advance\@pperedge by -1.2\baselineskip
        \ifdim\@pperedge>\pagetotal\global\everypar{}%
          \global\f@nishedtrue
        \else
          \s@tparshape                 % if we continue around the picture
        \fi
        \advance\@pperedge by 1.2\baselineskip
      \fi}
\def\s@tparshape{\t@mpdimen=-\pagetotal% minus upperedge of the picture
   \advance\t@mpdimen by\b@ttomedge    % plus bottomedge of the pictured
   \divide\t@mpdimen by\baselineskip   % divided by baselineskip gives the
   \helpc@unt=\t@mpdimen               %  number of shorter lines
   \advance \helpc@unt by 2            % rounding correction and one line more
   \sh@petoks=\expandafter{\the\helpc@unt\space}% \sh@petoks is for \parshape
   \t@mpdimeni=\lines@bove\baselineskip
   \t@mpdimen=\pagetotal
   \gdef\lines@bove{0}%              % in next paragraphs no lines are full
   \loop \ifdim\t@mpdimeni>0.999\baselineskip % counts the full lines
     \advance\t@mpdimen  by \baselineskip
     \advance\t@mpdimeni by-\baselineskip
     \sh@petoks=\expandafter{\the\sh@petoks 0pt \the\hsize}%
   \repeat
   \loop \ifdim\b@ttomedge>\t@mpdimen         % counts the shorter lines
     \advance\t@mpdimen by \baselineskip
     \sh@petoks=\expandafter{\the\sh@petoks \in@tdimen \l@nelength }%
   \repeat
   \sh@petoks=\expandafter
      {\the\sh@petoks 0pt \the\hsize}%        % and next lines are full again
   \expandafter\parshape\the\sh@petoks
   }

\descriptionmargins{2pt}
\picturemargins{15pt}{0pt}

\catcode`\@=\catcodeat        \let\catcodeat=\undefined